\documentclass[prd,preprint,eqsecnum,nofootinbib,amsmath,amssymb,
               tightenlines,dvips,floatfix]{revtex4}
\usepackage{graphicx}
\usepackage{bm}
\usepackage{amsfonts}
\usepackage{amssymb}
\newcommand{\mtrx}{\underline}

\usepackage{dsfont}

\def\b{{\bm b}}

\def\p{{\bm p}}
\def\q{{\bm q}}

\def\Nc{N_{\rm c}}

\def\mD{m_{\rm D}}
\def\alphas{\alpha_{\rm s}}

\def\tr{\operatorname{tr}}

\def\Tgen{{\mathbb T}}

\begin {document}



\title
    {
      Multi-particle potentials from light-like Wilson lines
      in quark-gluon plasmas:
      a generalized relation of in-medium splitting rates
      to jet-quenching parameters \boldmath$\hat q$
    }

\author{Peter Arnold}
\affiliation
    {%
    Department of Physics,
    University of Virginia,
    Charlottesville, Virginia 22904-4714, USA
    \medskip
    }%

\date {\today}

\begin {abstract}%
{%
  A powerful historical insight about the theory of in-medium
  showering in QCD backgrounds was that splitting rates can be related
  to a parameter $\hat q$ that characterizes the rate of transverse-momentum
  kicks to a high-energy particle from the medium.
  Another powerful insight was that $\hat q$ can be defined (with caveats)
  even when the medium is strongly coupled, using long, narrow
  Wilson loops whose
  two
  long edges are light-like Wilson lines.  The medium effects for the original
  calculations of in-medium splitting rates can be formulated in terms
  of 3-body imaginary-valued ``potentials'' that are defined with {\it three}
  long, light-like Wilson lines.  Corrections due
  to overlap of two consecutive splittings can be calculated using
  similarly defined 4-body potentials.
  I give a simple argument for how
  such $N$-body potentials can be determined
  in the appropriate limit
  just from knowledge of the values of
  $\hat q$ for different color representations.
  For $N > 3$, the $N$-body potentials have non-trivial color structure,
  which will complicate calculations
  of overlap corrections outside of the large-$\Nc$ or soft bremsstrahlung
  limits.
}%
\end {abstract}

\maketitle
\thispagestyle {empty}



\section{Overview and Result}
\label{sec:intro}

\begin {figure}[t]
\begin {center}
  \includegraphics[scale=0.4]{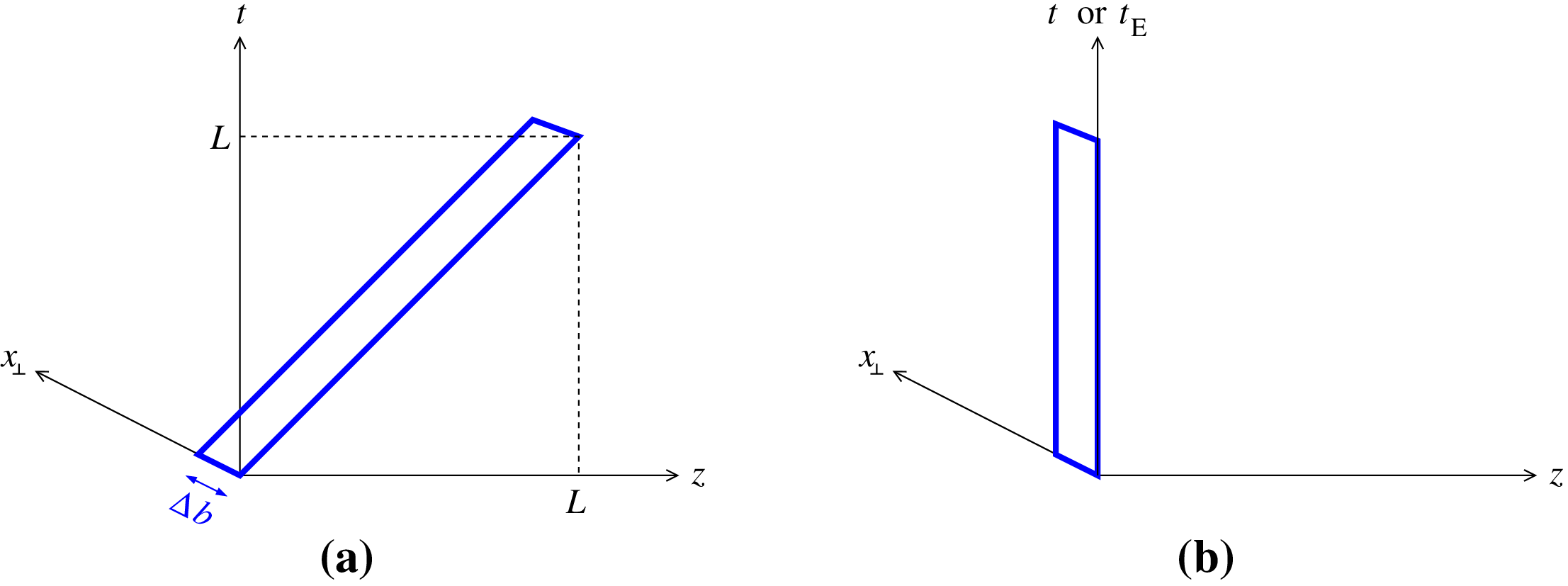}
  \caption{
     \label{fig:Wilson2}
     (a) shows a Wilson loop with light-like edges used to formally define
     $\hat q$ (subject to caveats mentioned in section \ref{sec:caveats}).
     $t$ is real (Minkowski) time.
     In contrast, (b) shows a Wilson loop for static color charges,
     where $t$ can be real or imaginary (Euclidean) time.
  }
\end {center}
\end {figure}

In theoretical studies of $p_\perp$-broadening and jet quenching of very
high-energy partons that travel through a quark-gluon plasma, a very
important parameter describing scattering of the parton from the medium
is known as $\hat q$.  Physically, it is the proportionality constant
in the relation $\langle Q_\perp^2 \rangle = \hat q \, \Delta z$, where
$\langle Q_\perp^2 \rangle$ is the typical squared transverse momentum
(transverse to the parton's initial direction of motion) that the parton picks
up after traveling a distance $\Delta z$ through the medium, in
the limit that $\Delta z$ is large compared to characteristic scales
of the medium such as mean-free paths for collisions.
As I briefly review later,
it has been known for some time \cite{LRW1,LRW2}
that $\hat q$ can also be formally
defined (with important caveats) in terms of a kind of ``potential energy''
$V(\Delta b)$ defined using a medium-averaged Wilson loop
having two long parallel
light-like sides separated by transverse distance $\Delta b$, as
shown in fig.\ \ref{fig:Wilson2}a.
This definition is similar to how potentials are
often defined for static charges using Wilson loops like
fig.\ \ref{fig:Wilson2}b.
The potential
$V(\Delta b)$ is extracted from the exponential dependence of the
Wilson loop on its length:
\begin {equation}
  \bigl\langle \tr[ P e^{ig\oint_C dx^\mu \, A_\mu} ] \bigr\rangle
  \approx e^{-i \, V(\Delta\b) \, L} ,
\label {eq:loop}
\end {equation}
where $P$ represents path-ordering in color space.%
\footnote{
  My $L$ corresponds to the $L^-/\sqrt2$ of refs.\ \cite{LRW1,LRW2}, and
  my transverse separation $\Delta b$ is what they call $L$.
  My characterization of the
  Wilson loop as defining a ``potential'' $V(\Delta\b)$ is not
  language specifically used by refs.\ \cite{LRW1,LRW2}.
}
Formally (again with important caveats),
one can show that $\hat q$ is the coefficient of a harmonic oscillator
approximation to this potential:
\begin {equation}
  V(\Delta b) \to -\tfrac{i}{4} \hat q \, (\Delta b)^2
  \quad \mbox{for small $\Delta b$,}
\label {eq:V2intro}
\end {equation}
which is equivalent to
\begin {equation}
  \bigl\langle \tr[ P e^{ig\oint_C dx^\mu \, A_\mu} ] \bigr\rangle
  \approx e^{-\hat q \, (\Delta b)^2 L/4}
  \quad \mbox{for small $\Delta b$.}
\end {equation}
The advantage of the Wilson loop
language is that it can be used as a tool for discussing
$\hat q$ in strongly (as well as weakly) coupled quark-gluon
plasmas.
In general, $\hat q$ depends on the color representation $R$ of
the high-energy particle.  For weakly-coupled plasmas, $\hat q$ is simply
proportional to the quadratic Casimir $C_R$ of that color representation,
but for strongly-coupled plasmas the
$\hat q$
for different color representations may not be so simply related.%
\footnote{
  The Casimir scaling $\hat q_R \propto C_R$
  holds through next-to-leading order in the strength of
  the coupling of the plasma \cite{SimonNLO},
  but
  this scaling need not hold exactly at all orders.
  For some examples of violation of Casimir scaling for Wilson loops in
  other contexts, see refs.\ \cite{Casimir1,Casimir2,Casimir3}.
}

\begin {figure}[t]
\begin {center}
  \includegraphics[scale=0.5]{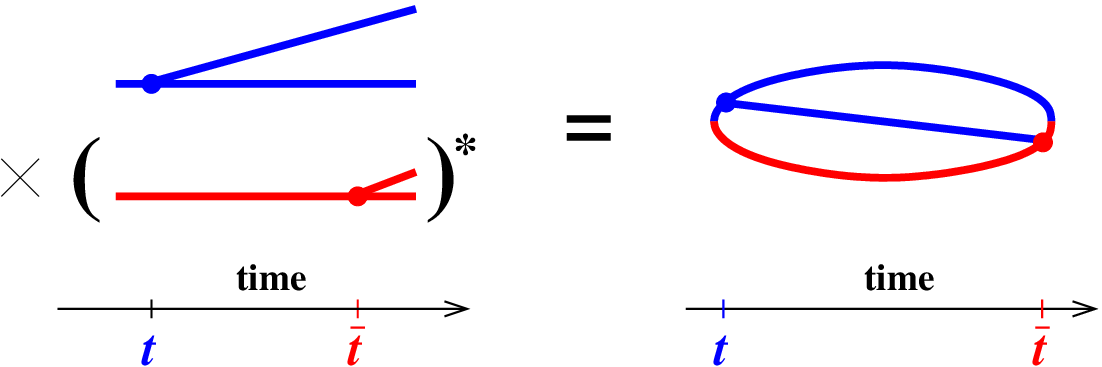}
  \caption{
     \label{fig:split}
     A contribution to the {\it rate}
     for single splitting of a high-energy particle
     in the medium.  Only high-energy particles are shown explicitly;
     all lines are implicitly interacting with the medium, which is then
     averaged over.  In these diagrams, time runs from left to right,
     and the amplitude and conjugate amplitude are each implicitly
     integrated over the time of emission ($t$ and $\bar t$ respectively)
     to get the splitting probability.  On the right-hand side is a combined
     diagram showing the amplitude (blue) sewn together with the conjugate
     amplitude (red).
  }
\end {center}
\end {figure}

\begin {figure}[t]
\begin {center}
  \includegraphics[scale=0.5]{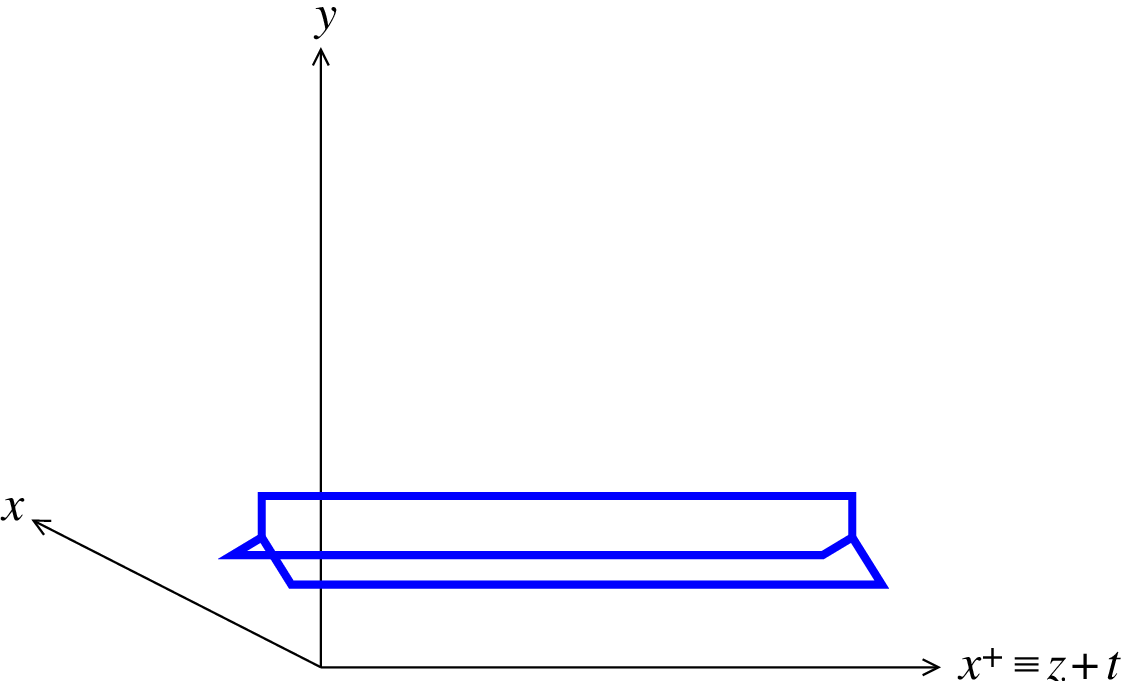}
  \caption{
     \label{fig:Wilson3}
     Like fig.\ \ref{fig:Wilson2}a but for a 3-body potential.
     Note that the axes are depicted differently
     than in fig.\ \ref{fig:Wilson2}:
     in order to be able to show both transverse spatial directions,
     the $t$ and $z$ axes have been collapsed to $x^+ = z+t$, with
     $x^- = 0$ everywhere.
     The 3-point vertices on the ends are chosen to form the
     color-neutral combination of the three particles.  For example,
     for the
     3-gluon potential, the Wilson lines would be adjoint representation
     and the 3-point vertices would each be proportional to the Lie algebra
     structure constants $f^{abc}$.
     Note: The constant transverse positions $(\b_1,\b_2,\b_3)$ of the three
     light-like Wilson lines can be anything; they need not be symmetrically
     arranged as in this picture.
  }
\end {center}
\end {figure}

Similar types of potentials arise
in calculations of splitting rates (bremsstrahlung or pair production)
in high-energy in-medium showers.  Splitting rate calculations are complicated
by the Landau-Pomeranchuk-Migdal (LPM) effect, which accounts for the
fact that high-energy particles can scatter from the medium many times
during the quantum mechanical duration, known as the formation
time, of a single splitting.
The QCD version of the LPM effect was originally worked out by
Baier et al.\ (BDMPS) \cite{BDMPS12,BDMPS3} and Zakharov \cite{Zakharov}.
Though they originally framed
their calculations in terms of a weakly-coupled picture of the medium,
the approach can be generalized to a strongly-coupled medium.
Consider the left-hand side of fig.\ \ref{fig:split},
which depicts an interference term that contributes to an in-medium splitting
rate.  Following Zakharov \cite{Zakharov}, one may sew together the diagrams
representing the amplitude and conjugate amplitude to form
the interference diagram on the right-hand side, which may now be formally
re-interpreted as the propagation of three particles through the
medium, where particles from the conjugate diagram (red)
are re-interpreted as their anti-particles.
Zakharov expresses the calculation of this interference in terms
of the time-evolution of the wave function of the transverse
positions of the three particles, which can be described by
a Schr\"odinger-like equation.
The potential term in that Schr\"odinger equation represents
medium-averaged effects
of interactions with the medium over time scales short compared to
the long formation time.  Over those time scales,
the transverse positions of the particles
can be treated as constant.
We could therefore identify this potential term as
the 3-particle potential
$V(\b_1,\b_2,\b_3)$ between light-like Wilson lines, such as depicted in
fig.\ \ref{fig:Wilson3}.  In the high-energy limit, the relevant
separations $\b_i{-}\b_j$ are small during the formation time because
splitting processes are nearly collinear, and
so one may make a harmonic oscillator approximation to
$V(\b_1,\b_2,\b_3)$.  As I will review,
there is a fairly simple argument \cite{2brem}%
\footnote{
  See in particular the discussion in Appendix A of ref.\ \cite{2brem}
  concerning eq.\ (2.21) of that paper.
  In actual splitting rate calculations, one can use symmetries
  (like BDMPS and Zakharov did) to reduce
  the work of solving the 3-body problem to solving an effective 1-body
  problem with a potential derived from (\ref{eq:V3}).
  For a description in the language used here, see
  sections II.E and III of ref.\ \cite{2brem}.
}
that, whenever a harmonic oscillator
approximation is relevant, it must necessarily have the form
\begin {multline}
   V(\b_1,\b_2,\b_3) =
      - \frac{i}8 \Bigl[
         (\hat q_1+\hat q_2-\hat q_3) (\b_2{-}\b_1)^2
\\
         + (\hat q_2+\hat q_3-\hat q_1) \,
             (\b_3{-}\b_2)^2
         + (\hat q_3+\hat q_1-\hat q_2) \,
             (\b_1{-}\b_3)^2
      \Bigr] ,
\label {eq:V3}
\end {multline}
where $\hat q_1$, $\hat q_2$, and $\hat q_3$ are the $\hat q$'s
for the color representations of the three high-energy particles
involved in the splitting process (the parent and the two daughters).
The argument for (\ref{eq:V3}) is simple in the sense that it does
not require any of the machinery of computing the LPM effect in QCD:
it just involves thinking through the constraints to any harmonic
3-body potential from the special cases where some of the separations
$\b_i{-}\b_j$ vanish.

There has been a variety of work on
the LPM effect in QCD studying the potentially
significant effects of what happens when two consecutive splittings
in an in-medium shower have overlapping formation times.
As I'll briefly review, such calculations generally require corresponding
4-particle potentials $V(\b_1,\b_2,\b_3,\b_4)$.
Including yet more particles would be needed
to study the simultaneous overlap of three or more
splittings.  To date, calculations of overlap effects
have made simplifying assumptions
such as soft emission limits \cite{Blaizot,Iancu,Wu}
or the large-$\Nc$ limit \cite{2brem,seq,dimreg,4point}.
In this paper, I take a first step toward
removing those assumptions by finding the generalization of
(\ref{eq:V3}) to four or more particles.  The result will be
\begin {equation}
   \mtrx{V}(\b_1,\b_2,\cdots,\b_N) =
      - \frac{i}8 \sum_{i>j}
         (\hat q_i+\hat q_j-\mtrx{\hat q}_{ij}) (\b_i{-}\b_j)^2 ,
\label {eq:VN}
\end {equation}
where underlines indicate an operator on the
space of color states of the $N$ particles.  The color structure
is necessary because
of the $\hat q_{ij}$ above, which refers to the {\it combined} color
representation of particles $i$ and $j$.  If both are gluons,
for example, the combined color representation could be any
irreducible representation $R$ in the SU(3) tensor product
$8\otimes8 = 1\oplus8\oplus8\oplus10\oplus\overline{10}\oplus27$,
and those representations generally have different values of $\hat q$.
The color operator $\mtrx{\hat q}_{ij}$ represents
use of the correct value of $\hat q$ in each color subspace.  Formally,
\begin {equation}
  \mtrx{\hat q}_{ij}
  =
  \sum_{R \in R_i\otimes R_j}
  \hat q_R \, \mtrx{\cal P}_{ij,R}
\end {equation}
where $\mtrx{\cal P}_{ij,R}$ is a projection operator, acting
on the $N$-particle color space, that selects
the subspace where particles $i$ and $j$ have combined
(irreducible) color representation $R$.

Since the particles described by the potential (\ref{eq:VN}) are
separated from each other (unless some $\b_i=\b_j$), readers may be concerned
about whether, in the general case, the ``combined color representation''
of any pair of particles $i$ and $j$ is a gauge-invariant
concept.  I'll later discuss the separation of scales
in this problem that addresses this point.

If one chooses a basis of the
$N$-particle color space where each basis element can be identified as
belonging to a particular irreducible combined color representation of
particles 1 and 2, then $\mtrx{\hat q}_{12}$ can be
represented as a diagonal matrix on the space of colors.  However,
in that basis, $\mtrx{\hat q}_{13}$ will generally {\it not}
be diagonal.  As a result, the potential $\mtrx{V}$ of
(\ref{eq:VN}) generically contains terms which mix different
possible color combinations of the $N$ particles.

The reason that this complication concerning color representations was
avoided in the 3-particle potential (\ref{eq:V3}) has to do with the
overall color state of the particles in the applications of interest.
Consider Zakharov's interpretation of the single splitting process,
depicted by the right-hand side of fig.\ \ref{fig:split}.  The total
color of the $N{=}3$ particles there is neutral.
The same is automatically true of
any gauge-invariant definition of a 3-body potential potential from
Wilson lines, such as in fig.\ \ref{fig:Wilson3}.  For $N{=}3$, overall
color neutrality means that the combined color representation of
particles 1 and 2 must be the conjugate of the color representation of
particle 3, so that $\mtrx{\hat q}_{12} = \hat q_3$ and similarly for
permutations.  This is how the $N$-body result (\ref{eq:VN}) reduces
to the simpler 3-body result (\ref{eq:V3}) for $N{=}3$.

\begin {figure}[t]
\begin {center}
  \includegraphics[scale=0.7]{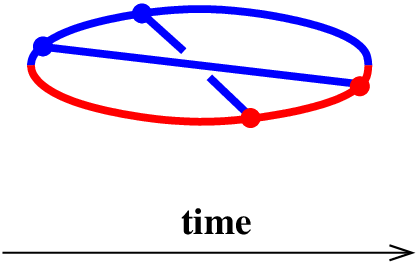}
  \caption{
     \label{fig:split2}
     Similar to the right-hand side of fig.\ \ref{fig:split} but
     for double splitting.
  }
\end {center}
\end {figure}

Fig.\ \ref{fig:split2} similarly shows an example
of an interference
term
for the process of double splitting, in a case relevant to computing
effects of overlapping formation times \cite{2brem}.
For part of the time evolution in this example,
there are $N{=}4$ particles, also forming an overall color singlet.
For $N{>}3$, this overall color neutrality does not
constrain the combined color representations of each pair of particles
to a unique irreducible representation, and so the matrix structure
of the potential $\mtrx{V}$ in color space is unavoidable.

In general,
$N$-body potentials for overall color singlets are a tool for
consolidating all interactions with the medium
that occur over time scales small compared to the time
scale of the splitting
processes shown in figs.\ \ref{fig:split} and \ref{fig:split2},
i.e.\ on time scales small compared to formation times.


\subsection*{Outline}

Before proceeding to a general argument for the $N$-body potential
(\ref{eq:VN}),
it may help motivate the color structure of the result
to first discuss the special case of weakly-coupled plasmas
in section \ref{sec:weak}.
There I start with a brief review of the relation of $\hat q$
to the 2-body potential, and then generalize to a discussion
of the $N$-body potential.
That section is not necessary, though, for readers wishing
to quickly cut to the chase and see the
general argument for the result (\ref{eq:VN}), which is given
in section \ref{sec:general}.

In section \ref{sec:caveats}, I list many of the caveats that I have
been sweeping under the rug concerning the definition and meaning of
$\hat q$ and harmonic oscillator approximations to the potential.
I also explain the hierarchy of time scales that makes it sensible
to discuss a potential with non-trivial color structure.
Section \ref{sec:conclusion} offers a brief recap and
conclusion.


\section {The special case of weakly-coupled plasmas}
\label{sec:weak}


\subsection {2-body potential and \boldmath$\hat q$}

First, I start with a brief review of the physics behind the
2-body potential.  It is useful to forget about Wilson loops for
a moment and first review the probabilistic evolution of the
$\p_\perp$ of a high-energy particle receiving random transverse
momentum kicks as it crosses the plasma.%
\footnote{
  The review in section \ref{sec:ptreview} is a generalization of
  BDMPS's eqs. (2.8--2.12) and (3.1) of ref.\ \cite{BDMPS3}.
  By casting the derivation
  in terms of $d\Gamma_{\rm el}/d^2 q_\perp$ \cite{simple} instead of
  BDMPS's $V(Q^2)$, the review here
  avoids BDMPS's model assumption that the medium can be treated as
  a collection of static scattering centers.
}


\subsubsection {Evolution of transverse momentum}
\label {sec:ptreview}

For a high-energy particle, we can follow the evolution
of the classical probability distribution $\rho(\p_\perp)$ of its transverse
momentum using the equation
\begin {equation}
  \frac{\partial\rho(\p_\perp,t)}{\partial t}
  =
  -
  \int d^2 q_\perp \>
  \frac{d\Gamma_{\rm el}}{d^2 q_\perp} \,
  \bigl[
    \rho(\p_\perp,t) - \rho(\p_\perp{-}\q_\perp,t)
  \bigr] ,
\label {eq:FP}
\end {equation}
where $d\Gamma_{\rm el}/d^2q_\perp$ is the differential rate for the
high-energy particle to pick up transverse momentum $\q_\perp$ from the
medium.
The first term on the right-hand side of (\ref{eq:FP})
is a loss term,
corresponding to the chance for momentum $\p_\perp$ to be scattered
to some other other momentum; the second term is a gain term,
corresponding to some other momentum
scattering to become $\p_\perp$.
The equation can be solved by Fourier transforming from $\p_\perp$
to transverse position space:
\begin {equation}
  \frac{\partial\rho(\b,t)}{\partial t}
  =
  - \Delta\Gamma_{\rm el}(\b) \,
  \rho(\b) ,
\label {eq:FPb}
\end {equation}
where
\begin {equation}
  \Delta\Gamma_{\rm el}(\b) \equiv \Gamma_{\rm el}(0) - \Gamma_{\rm el}(\b)
  =
  \int d^2 q_\perp
  \frac{d\Gamma_{\rm el}}{d^2 q_\perp} \, (1 - e^{i\b\cdot\q_\perp})
\label {eq:DeltaGamma}
\end {equation}
and
\begin {equation}
  \Gamma_{\rm el}(\b) \equiv
  \int d^2 q_\perp
  \frac{d\Gamma_{\rm el}}{d^2 q_\perp} \, e^{i\b\cdot\q_\perp} .
\label {eq:Gammab}
\end {equation}
Multiplying both sides of (\ref{eq:FPb}) by $i$,
the evolution equation may be formally recast
as a Schr\"odinger-like equation
\begin {equation}
  i \partial_t \rho(\b,t)
  =
  V(\b) \, \rho(\b,t)
\end {equation}
with no kinetic term and with imaginary-valued potential energy
\begin {equation}
  V(\b) = -i \,\Delta\Gamma_{\rm el}(\b) .
\label {eq:VdG}
\end {equation}
The corresponding solution is
\begin {equation}
  \rho(\b,t) = e^{-i V(\b) t} \rho(\b,0)
  = e^{-\Delta\Gamma_{\rm el}(\b)\,t} \rho(\b,0) .
\end {equation}


\subsubsection {The same potential from a Wilson loop}

\begin {figure}[t]
\begin {center}
  \includegraphics[scale=0.4]{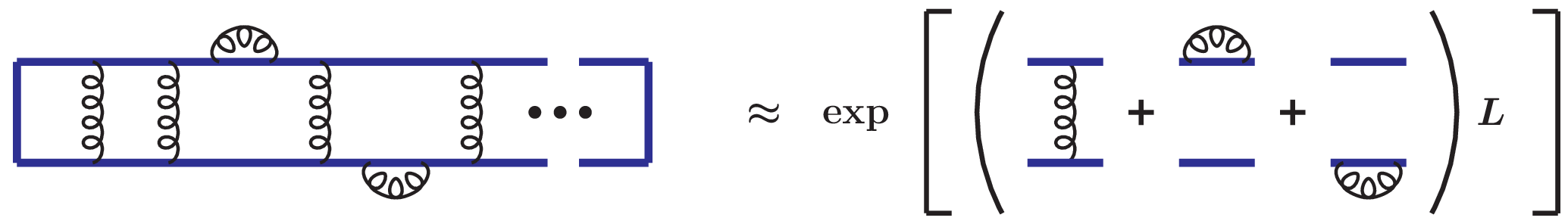}
  \caption{
     \label{fig:Wilson2weak}
     The Wilson loop in the weak-coupling limit.  Here, the gluon
     lines represent 2-point correlators of the gauge field in the
     background of the medium.
     The particular graph on the left-hand
     side is just an example. The important point is that, in the
     weak-coupling limit, localized non-overlapping 2-point correlations
     dominate and exponentiate as shown on the right-hand side.
     Each 2-point correlation (depicted by a gluon line) is implicity resummed
     with in-medium self-energy insertions.
  }
\end {center}
\end {figure}

At leading order in the weak-coupling limit (with resummation of
in-medium self energies), this same physics arises
from light-like Wilson loops via 2-point correlators of interactions of
the Wilson lines with background gauge fields, as depicted in
fig.\ \ref{fig:Wilson2weak}.  Those correlators
between Wilson lines can be shown to correspond to
$\Gamma_{\rm el}(\b_i{-}\b_j)$.  [For the sake of completeness,
I review this in the appendix, but the details will not be important.]
Each self-energy loop on a Wilson line is additionally associated
with a factor%
\footnote{
  When integrating over
  the relative time $\Delta x^+$ between the two endpoints of a
  2-point correlator in fig.\ \ref{fig:Wilson2weak}, the integral
  is $\int_{-\infty}^{\infty} d(\Delta x^+)$ for correlators that
  span two different Wilson lines ($i{\not=}j$) but, to avoid
  double counting of a given loop, is instead
  $\int_0^\infty d(\Delta x^+) = \tfrac12 \int_{-\infty}^{\infty} d(\Delta x^+)$
  for self-energy corrections to a single Wilson line ($i{=}j$).
  The relative minus sign for self-energies compared to correlations
  between different Wilson lines is because the color path-ordering
  of the two Wilson lines are in opposite directions.
  (Equivalently, if one wants to view both lines as running forward
  in time, then it is because the color charges $\Tgen_1$ and
  $\Tgen_2$ of
  the two lines must be opposite by overall color neutrality
  $\Tgen_1+\Tgen_2=0$ of the Wilson loop.  See section \ref{sec:Nbodyweak}.)
}
of $-\tfrac12$,
and so the exponent in
fig.\ \ref{fig:Wilson2weak} is $-i\,V(\Delta\b)\,L$ with
\begin {equation}
  V(\Delta\b) = -i[\Gamma_{\rm el}(0) - \Gamma_{\rm el}(\Delta\b)],
\label {eq:Wilson2}
\end {equation}
reproducing the potential (\ref{eq:VdG}).

In this weak-coupling limit, the only dependence on the color representation
of the high-energy particle is that $d\Gamma_{\rm el}$ is proportional
to $g^2 C_R$, where $C_R$ is the quadratic Casimir of that representation.
It will be useful in the remaining discussion of the weakly-coupled limit
to explicitly pull out this Casimir and write
\begin {equation}
  \Gamma_{\rm el} = C_R \bar\Gamma_{\rm el} ,
  \qquad
  V(\Delta\b) = C_R \bar V(\Delta\b) ,
  \qquad
  \hat q = C_R \hat{\bar q} ,
  \qquad
  \mbox{etc.},
\end {equation}
where the barred quantities do not (in weak coupling) depend on the
color representation of the high-energy particle.


\subsubsection {Relation to $\hat q$}

Formally expanding (\ref{eq:DeltaGamma}) in powers of $\b$,
the small-$b$ limit is
\begin {equation}
  V(\b) \simeq
    -\tfrac{i}{4} b^2 \int d^2 q_\perp \>
    \frac{d\Gamma_{\rm el}}{d^2 q_\perp} \, q_\perp^2 .
\label {eq:Vqhat2a}
\end {equation}
This is related to $\hat q$:
The rate at which the 
$p_\perp^2 = |\q_{\perp1}+\q_{\perp2}+\cdots|^2$ of a hard
particle would increase with time
from an initial $p_\perp^2 \equiv 0$,
by a sequence of random kicks $\q_\perp$ from the medium, is given by
\begin {equation}
  \hat q = \int d^2 q_\perp \> \frac{d\Gamma_{\rm el}}{d^2 q_\perp} \, q_\perp^2 .
\label {eq:qhat}
\end {equation}
So (\ref{eq:Vqhat2a}) is
\begin {equation}
  V(\b) \simeq
    -\tfrac{i}{4} \hat q b^2 .
\label {eq:Vqhat2}
\end {equation}


\subsection {The $N$-body potential}
\label {sec:Nbodyweak}

\begin {figure}[t]
\begin {center}
  \includegraphics[scale=0.4]{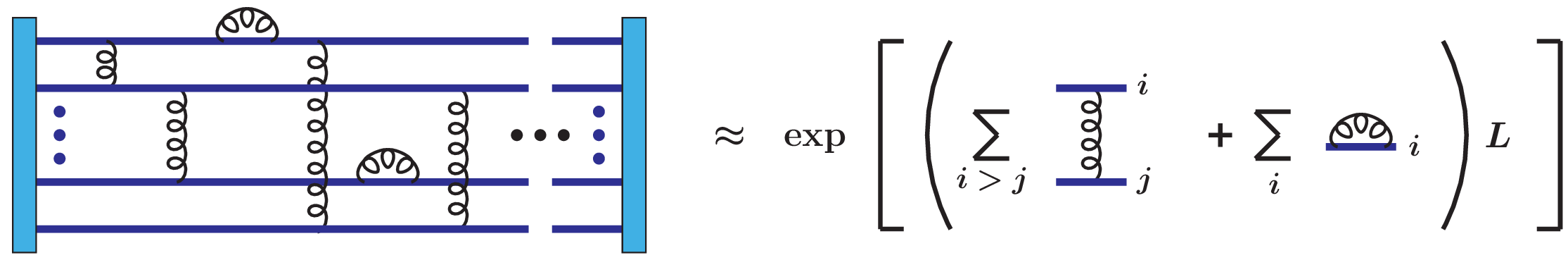}
  \caption{
     \label{fig:WilsonNweak}
     Like fig.\ \ref{fig:Wilson2weak} but for an $N$-body potential.
     The blue rectangles above indicate some contraction of the light-like
     Wilson
     lines to form a gauge-invariant quantity (and so an overall color-neutral
     state of the $N$ high-energy particles represented by the Wilson lines).
  }
\end {center}
\end {figure}

Fig.\ \ref{fig:WilsonNweak} shows a similar set of correlators for an
$N$-body potential.  The charge with which
Wilson line $i$ interacts with a background gauge field of
adjoint color $a$ is
$g \mtrx{\Tgen}_i^a$, where the $\mtrx{\Tgen}_i^a$ are color generators
in the color representation of particle $i$, acting on the color
space of that particular particle.
In the weak-coupling limit,
fig.\ \ref{fig:WilsonNweak} shows that
the $N$-body potential then decomposes
into 2-body correlators as
\begin {equation}
   V(\b_1,\b_2,\cdots,\b_N) =
      -i \biggl\{
        \tfrac12
        \sum_i \mtrx{\Tgen}_i^2 \, \bar\Gamma_{\rm el}(0)
        + \sum_{i>j} \mtrx{\Tgen}_i\cdot \mtrx{\Tgen}_j
             \bar\Gamma_{\rm el}(\b_i{-}\b_j)
       \biggr\} ,
\label {eq:WilsonN0}
\end {equation}
which is the generalization of (\ref{eq:Wilson2}).
Above, $\mtrx{\Tgen}_i \cdot \mtrx{\Tgen}_j$ represents the sum
$\mtrx{\Tgen}_i^a \mtrx{\Tgen}_j^a$ over $a$.
Since for the applications of interest
the collection of $N$ particles is overall color neutral, we
can subtract
$0 = -\tfrac{i}{2}(\mtrx{\Tgen}_1+\mtrx{\Tgen}_2+\cdots\mtrx{\Tgen}_N)^2
     \,\bar\Gamma_{\rm el}(0)$
from (\ref{eq:WilsonN0}) to rewrite it in the form
\begin {equation}
   V(\b_1,\b_2,\cdots,\b_N) =
      i \sum_{i>j} \mtrx{\Tgen}_i\cdot \mtrx{\Tgen}_j
             \,\Delta\bar\Gamma_{\rm el}(\b_i{-}\b_j)
   =
      - \sum_{i>j} \mtrx{\Tgen}_i\cdot \mtrx{\Tgen}_j
             \,\bar V_{(2)}(\b_i{-}\b_j) ,
\end {equation}
where $\bar V_{(2)}$ is the universal (in weak coupling) two-body potential
when the color generators are factored out.
For small transverse separations, (\ref{eq:Vqhat2}) then gives
\begin {equation}
   V(\b_1,\b_2,\cdots,\b_N) \simeq
     \tfrac{i}{4}
     \sum_{i>j} \mtrx{\Tgen}_i\cdot \mtrx{\Tgen}_j
           \hat{\bar q} \, (\b_i{-}\b_j)^2 .
\label {eq:VqhatN0}
\end {equation}
Rewriting
\begin {equation}
  \mtrx{\Tgen}_i\cdot \mtrx{\Tgen}_j
  = \tfrac12 \bigl[
       (\mtrx{\Tgen}_i + \mtrx{\Tgen}_j)^2 - \mtrx{\Tgen}_i^2 - \mtrx{\Tgen}_j^2 
    \bigr]
  = \tfrac12 \bigl[ (\mtrx{\Tgen}_i + \mtrx{\Tgen}_j)^2 - C_i - C_j \bigr] ,
\end {equation}
and remembering that
${\hat q}_R = C_R \hat{\bar q} = \mtrx{\Tgen}_R^2 \hat{\bar q}$
(in weak coupling),
the potential (\ref{eq:VqhatN0}) can be recast as
\begin {equation}
   \mtrx{V}(\b_1,\b_2,\cdots,\b_N) =
      - \frac{i}8 \sum_{i>j}
         (\hat q_i+\hat q_j-\mtrx{\hat q}_{ij}) (\b_i{-}\b_j)^2 .
\end {equation}
This demonstrates the result (\ref{eq:VN}) in the special case of
weak coupling.

\section {General argument for any strength coupling}
\label {sec:general}

Generically, if a function $V(\b_1,\cdots,\b_N$)
is translationally and rotationally invariant (here in the transverse
plane) and approaches a smooth limit as all the $\b_i$ become
coincident, we may suspect that it can be Taylor expanded in
powers of $\b_i{-}\b_j$ to obtain a harmonic oscillator approximation
in the limit of small separations.
Here, we now give a simple argument why, if there {\it is} such a
harmonic oscillator approximation, then its form is fixed
as (\ref{eq:VN}), provided the $N$ particles in total form a color
singlet.

Start by assuming a generic harmonic oscillator form that is
(transversely) translation invariant:
\begin {equation}
   \mtrx{V}(\b_1,\b_2,\cdots,\b_N)
   = \sum_{ijkl} \mtrx{c}_{ijkl} (\b_i{-}\b_j)\cdot(\b_k{-}\b_l)
\label {eq:harmonic1}
\end {equation}
for some constants $\mtrx{c}_{ijkl}$.
Those constants must be allowed to be color operators, since
we have already seen that's necessary in the special case (\ref{eq:VqhatN0}) of
weak coupling.  Using the algebraic identity
\begin {equation}
  (\b_i{-}\b_j)\cdot(\b_k{-}\b_l) =
  \tfrac12 \bigl[
     (\b_i{-}\b_l)^2 + (\b_j{-}\b_k)^2 - (\b_i{-}\b_k)^2 - (\b_j-\b_l)^2
  \bigr] ,
\end {equation}
any potential of the form (\ref{eq:harmonic1}) can be recast into the
form
\begin {equation}
   \mtrx{V}(\b_1,\b_2,\cdots,\b_N)
   = \sum_{i>j} \mtrx{a}_{ij} (\b_i{-}\b_j)^2
\label {eq:harmonic2}
\end {equation}
for some constants $\mtrx{a}_{ij}$.


\subsection {The 3-particle case}
\label {sec:V3}

Before continuing with the $N$-particle case, it will be helpful to
review the logic of the 3-particle case:
\begin {equation}
   \mtrx{V}(\b_1,\b_2,\b_3) =
   \mtrx{a}_{21} (\b_2-\b_1)^2
   + \mtrx{a}_{32} (\b_3-\b_2)^2
   + \mtrx{a}_{13} (\b_1-\b_3)^2 .
\end {equation}
In the special case $\b_1 = \b_2$, this gives
\begin {equation}
   \mtrx{V}(\b_1,\b_1,\b_3) = (\mtrx{a}_{32}+\mtrx{a}_{13})(\b_3-\b_1)^2 .
\label {eq:V3special}
\end {equation}
However, in this case we have color charge $\Tgen_3$ at $\b_3$ and,
since $\Tgen_1+\Tgen_2+\Tgen_3=0$, total
color charge $\Tgen_1+\Tgen_2=-\Tgen_3$ at $\b_1=\b_2$.
This is then effectively a
2-body problem, in the color representation of particle 3,
with separation $\Delta\b = \b_3-\b_1$.  That means that
\begin {equation}
   \mtrx{V}(\b_1,\b_1,\b_3) = - \frac{i}{4} \hat q_3 (\b_3 - \b_1)^2
\end {equation}
in quadratic approximation, as in (\ref{eq:Vqhat2}).
Combined with (\ref{eq:V3special}),
this gives a constraint
$\mtrx{a}_{32}+\mtrx{a}_{13} = -\tfrac{i}{4} \hat q_3$
on the values of the coefficients
$\mtrx{a}_{ij}$.  Permuting the particle labels in this argument
then provides three constraints on the three unknown coefficients
$\mtrx{a}_{ij}$, which then uniquely determine the 3-body potential
to be (\ref{eq:V3}) in harmonic approximation.  For what follows,
I will find it more useful to write that 3-body potential in the
form (\ref{eq:VN}) that I will use for the $N$-body potential:
\begin {equation}
   \mtrx{V}(\b_1,\b_2,\b_3) =
      - \frac{i}8 \Bigl[
          (\hat q_2+\hat q_1-\mtrx{\hat q}_{21}) (\b_2{-}\b_1)^2
          + (\hat q_3+\hat q_2-\mtrx{\hat q}_{32}) (\b_3{-}\b_2)^2
          + (\hat q_1+\hat q_3-\mtrx{\hat q}_{13}) (\b_1{-}\b_3)^2
       \Bigr] .
\label {eq:V3alt}
\end {equation}
As previously noted, in the 3-body case, $\mtrx{\hat q}_{21}$ is
the same as $\hat q_3$, and so forth.


\subsection {The \boldmath$N$-particle case}

Now return to the generic form (\ref{eq:harmonic2}) for a harmonic
oscillator approximation in the $N$-body case.
Consider now the special case where we put all the particles
but the first two at the same position
$\b_3 = \b_4 = \cdots = \b_N$, so that (\ref{eq:harmonic2}) gives
\begin {equation}
  V(\b_1,\b_2,\b_3,\b_3,\cdots,\b_3) =
     \mtrx{a}_{21} (\b_2-\b_1)^2
   + \bigl(\sum_{j\ge 3} \mtrx{a}_{j1}\bigr) (\b_3-\b_1)^2
   + \bigl(\sum_{j\ge 3} \mtrx{a}_{j2}\bigr) (\b_3-\b_2)^2 .
\label {eq:VNspecial}
\end {equation}
On the other hand, $\b_3 = \b_4 = \cdots = \b_N$
is effectively a 3-body problem where one
particle has color generator 
$\mtrx{\Tgen}_3+\cdots\mtrx{\Tgen}_N$.
So (\ref{eq:VNspecial}) must be the same as
the 3-body potential (\ref{eq:V3alt}) with the
replacement $\mtrx{\Tgen}_3 \to \mtrx{\Tgen}_3+\cdots\mtrx{\Tgen}_N$
in the latter.
Focusing on the $(\b_2-\b_1)^2$ term for simplicity, that identification
requires
\begin {equation}
   \mtrx{a}_{21} =
   - \tfrac{i}8 (\hat q_2+\hat q_1-\mtrx{\hat q}_{21}) .
\end {equation}
There was nothing special about which of the $N$ particles we
labeled as $1$ and $2$ in this argument, so generally
\begin {equation}
   \mtrx{a}_{ij} =
   - \tfrac{i}{8} (\hat q_i+\hat q_j-\mtrx{\hat q}_{ij}) .
\end {equation}
Substitution into (\ref{eq:harmonic2}) then gives the final
result (\ref{eq:VN}) of this paper for the $N$-body potential
in harmonic approximation.


\section {Caveats and Clarifications}
\label {sec:caveats}

\subsection {Caveats for \boldmath$\hat q$ approximation}

\subsubsection {Logarithmic dependence}

The 2-body potential, which I've called
$V(\Delta b)$, is not precisely quadratic (\ref{eq:V2intro}) in
the small-$\Delta b$ limit.  Instead, the coefficient $\hat q$
effectively depends logarithmically on $\Delta b$.
One type of possible logarithmic dependence can be seen in
leading-order calculations of $\hat q$, where
$d\Gamma_{\rm el}/d^2 q_\perp \propto \alpha^2/q_\perp^4$ for
large $q_\perp$\thinspace: if one does not
account for any running of the coupling constant $\alphas$,
the leading-order result for $\hat q(\Delta b)$ blows up logarithmically
as $\Delta b \to 0$, so that $\hat q(0)$ as defined by (\ref{eq:qhat})
is ultraviolet (UV) divergent.  If one instead uses the running coupling
$\alphas(q_\perp)$ when calculating (\ref{eq:qhat}), the log dependence
of $\hat q(\Delta b)$ cuts off when $\alphas(1/\Delta b) \ll \alphas(\mD)$
[where $\mD$ is a plasma scale, representing the Debye mass],
and the leading-order result for $\hat q(0)$ is finite.%
\footnote{
  See, for example, section VI.B of ref.\ \cite{ArnoldDogan},
  which combines earlier
  observations of refs.\ \cite{BDMPS3} and \cite{Peshier}.
}
But there remains {\it other} log dependence that cannot be
seen at leading order \cite{Wu0}, which I will later review in section
\ref{sec:dblLog} below.

The arguments in this paper (like most any application of
the $\hat q$ approximation) rely on logarithmic
dependence of $\hat q$ being mild enough that one can simply approximate
the coefficient $\hat q$ by some fixed effective value relevant to
the scale of a particular application.


\subsubsection {Applies to typical events}

Another issue with the $\hat q$ approximation is that $\hat q$ only
determines the transverse momentum transfer for {\it typical}\/ multiple
scattering events.  Because of large-$q_\perp$ tails to the probability
distribution for momentum transfer in Coulomb scattering, there
are also rarer events with scattering by larger-than-typical angles.
Depending on the situation and what average quantity one is interested
in calculating, atypical events can sometime dominate averages.%
\footnote{
  See BDMPS, section 3.1 of ref.\ \cite{BDMPS3},
  and Zakharov \cite{ZakharovHO}.
  Some further discussion is given in ref.\ \cite{HO}.
}


\subsubsection {The limit of light-like Wilson lines}

Consider Wilson lines corresponding to particles with velocity $v$.
The light-like Wilson lines used in Wilson loops like
fig.\ \ref{fig:Wilson2}a
to define $\hat q$ [or more generally $\hat q(\Delta b)$] correspond to
the limiting case $v{=}1$.  There have been some confusing
subtleties in the literature on how to approach this limit---discussion
which has been in the context of calculations of $\hat q$
in QCD-like theories with gravity duals and which is also related to issues
of regularizing UV divergences associated with Wilson lines.
Here, I want to make a few simple observations about the $v{=}1$
limit in the general context of gauge theories, and then I will
draw some parallel to the issues in gauge-gravity duality calculations
at the end.

Physically, the test
particles represented by the long sides of the Wilson loop
should have $v < 1$, and so the light-like limit represents approaching
$v{=}1$ from below.  A natural impulse is to hope that if the limit
makes sense, then one should also be able to approach $v{=}1$ from
above---that is, using Wilson loops whose long sides are
(slightly)%
\footnote{
  Here ``slightly space-like'' means as considered in the plasma rest frame.
}
space-like rather than (slightly) time-like.  If so, then there are simple,
direct arguments that the Wilson loops, and the potentials $V(b)$
defined by them, have the following very nice and relevant properties.

First, time-ordering prescriptions do not matter for gauge fields
sourced by space-like Wilson loops.
In (all-orders) perturbative language, for example,
consider a correlator
\begin {equation}
   \langle A_{\mu_1}^{a_1}(x_1) \cdots A_{\mu_n}^{a_n}(x_n) \rangle
\label{eq:Acorrelator}
\end {equation}
of gauge fields located at $n$ different points $x_i$ on
the spatial Wilson loop.
Because the $x_i$ are then all space-like separated from each other,
any operators at different $x_i$ must commute because of causality.
So the {\it ordering} of the fields in the correlator (\ref{eq:Acorrelator})
is irrelevant.  That means we will get the same answer if we use
time-ordered correlators, anti-time-ordered correlators, Wightman
(un-ordered) correlators, Schwinger-Keldysh, or whatever.%
\footnote{
  Non-abelian Wilson loops still have very important {\it color} ordering,
  represented by path ordering in color space of the exponential
  $P \exp(i g \oint_C A^\mu \> dx_\mu)$ defining the Wilson loop.
  This only affects how the color indices $a_i$ in (\ref{eq:Acorrelator})
  will be contracted.
}

This is a significant property which has been implicitly
assumed in
applications of leading-order BDMPS-Z splitting
rates to strongly-coupled quark-gluon plasmas.
If time-ordering matters, then there is no general reason to think that
correlations between medium interactions of the two blue lines
(representing a pair of particles in the amplitude) in fig.\ \ref{fig:split}
is the same as the correlations between medium interactions of a blue and
a red line (representing one particle in the amplitude and one in the
conjugate amplitude).  The derivation given in section \ref{sec:V3}
for the $\hat q$ approximation to the 3-body potential (\ref{eq:V3})
would then be invalid.  That would in turn cast doubt on the
applicability of the standard BDMPS-Z formula for $g \to gg$ splitting
in $\hat q$ approximation in the case of strongly-coupled plasmas
(outside of the soft bremsstrahlung approximation, at least).

I should note that previous authors%
\footnote{
  See section VI of ref.\ \cite{D'Eramo}.
  (For a discussion of gauge invariance in the case of that ordering
  prescription, see also ref.\ \cite{Benzke}.)
}
have made a point that $\hat q$
should be defined with Schwinger-Keldysh ordering,
which is a natural choice when thinking of the meaning of $\hat q$
in terms of the rate for $\p_\perp$ broadening.
But if this ordering distinction is actually
important in the light-like limit,
we must then face the just-discussed difficulty
when using $\hat q$ for BDMPS-Z splitting rates.

A related nice property, also easy to derive if one may approach
light-like Wilson loops $v{=}1$ from the limit of space-like loops
$v>1$, is that expectations of Wilson loops are real-valued, and so
the potential $V(b)$ is pure imaginary as I assumed earlier.
Correspondingly $\hat q$ defined by (\ref{eq:V2intro}) is then
real-valued.  One way to see this is that, if time-ordering prescriptions
don't matter for space-like loops, then taking the complex conjugation
of the Wilson loop $\tr[P \exp(i\oint_C A^\mu \> dx_\mu)]$ is
equivalent to flipping the direction of integration around the loop,
which is equivalent to rotating the loop by 180 degrees in the transverse
plane.  By rotational invariance, the result for the Wilson
loop must therefore equal its complex conjugate and so is real.

In discussing details of their gauge-gravity duality calculation,
Liu, Rajagopal, and Wiedemann \cite{LRW2}
do not consider $v > 1$.
They characterize their calculation as approaching light-like
Wilson loops from the (physically motivated) $v < 1$ side but note
that they have to take the $v{=}1$ limit {\it before} they remove their
UV regulator, which corresponds to a tiny displacement of their string
endpoints into the fifth dimension.%
\footnote{
  Specifically, see the discussions of orders of limits in
  section 3 of ref.\ \cite{LRW2} and especially the conclusion of
  their section 3.3.
}
However, with their UV regulator in place, in the gravity description the
endpoints of their strings are forced to move faster than the local speed of
light (as noted in ref.\ \cite{LRW3}),
which is a reflection of the fact that the string worldsheets they
find are purely space-like in the order of limits that they take.
Their result is the same as they would get if they took their limit
from the $v > 1$ side.

If the limits of approaching $v{=}1$ from above and below are not the
same for all applications of Wilson loop potentials to splitting rates,
and furthermore if gauge field ordering prescriptions {\it do} matter in
the light-like limit, then one will have to figure out how to appropriately
adjust both BDMPS-Z splitting rates and the more generalized discussion
of this paper.


\subsection {Time and distance scales for color dynamics}

Earlier, I postponed discussing what it means to talk about
the joint color representation $\mtrx{\Tgen}_i + \mtrx{\Tgen}_j$ of two
spatially-separated particles $i$ and $j$.


\subsubsection {An over-simplified argument}

For this purpose, it is important to realize that the background
fields of the plasma will have characteristic wavelengths and correlation
lengths set by plasma scales (e.g.\ $1/T$ or $1/gT$ or $1/g^2 T$, etc.,
where $T$ represents the temperature).
But the $\hat q$ approximation relevant to high-energy particles
($E \gg T$) corresponds
to $\Delta b$ small compared to plasma scales.
That means that the different high-energy particles
represented by the light-like Wilson lines in
the 2-body or $N$-body potentials are so close to each other
than, to first approximation, they will experience
the {\it same} background gauge field
[$A_{\rm plasma}(\b_i,z{=}t) \simeq A_{\rm plasma}(\b_j,z{=}t)$].%
\footnote{
  Technically, pursuing the argument in the language of gauge
  fields $A$ requires
  insisting that the choice of gauge respects the separation of
  physics scales, so that plasma gauge fields are smooth on small scales
  (i.e. $\ll 1/T$). I assume that and proceed.
}
If two
slightly-separated Wilson lines $i$ and $j$ are experiencing
identical gauge fields as they move through light-cone time $x^+$,
then their color charges $\mtrx{\Tgen}_i$ and $\mtrx{\Tgen}_j$ will rotate
the same way, and so the sum $\mtrx{\Tgen}_i+\mtrx{\Tgen}_j$ will only
experience an overall color rotation.  In this approximation,
if one starts in a given color subspace corresponding to some
irreducible representation of $\mtrx{\Tgen}_i+\mtrx{\Tgen}_j$, then one
remains in that irreducible representation.

However, no matter how close the $\b_i$ are, there will be small differences
in the background color fields experienced by the different particles, and
these will slowly accumulate over time.%
\footnote{
  For a discussion of the language of color decoherence in the context of
  in-medium radiative processes, see, for example,
  ref.\ \cite{YacineColorSplit}.
  The parametric estimates [such as their (5.2)] are different because they
  consider the case of lines at some angle
  $\theta_{q\bar q}$ to each other,
  instead of the parallel lines with fixed separation $\b_{ij}$ relevant
  to defining the potential $V$ here.
}
The individual colors will decohere
over a time known as the color decoherence time $t_{\rm decohere}$,
which is parametrically of size
\begin {equation}
   t_{\rm decohere} \sim \frac{1}{\hat q \, (\Delta b)^2} \,.
\label {eq:tdecohere}
\end {equation}
This is just the time scale $L$ for which the values $e^{-i V L}$
of the Wilson loops discussed
in this paper [e.g.\ (\ref{eq:loop}) with (\ref{eq:V2intro})]
first become significantly different from one.
It is also the same time scale as formation times in applications to
in-medium splitting rates.
The important point is that, in the limit of small $\Delta b$, the
color decoherence time (\ref{eq:tdecohere})
is parametrically large compared to all plasma scales, which
means that the irreducible color representations of
$\mtrx{\Tgen}_i+\mtrx{\Tgen}_j$ only mix {\it slowly} compared to the
correlation length $\xi$ of the medium.

\begin {figure}[t]
\begin {center}
  \includegraphics[scale=0.5]{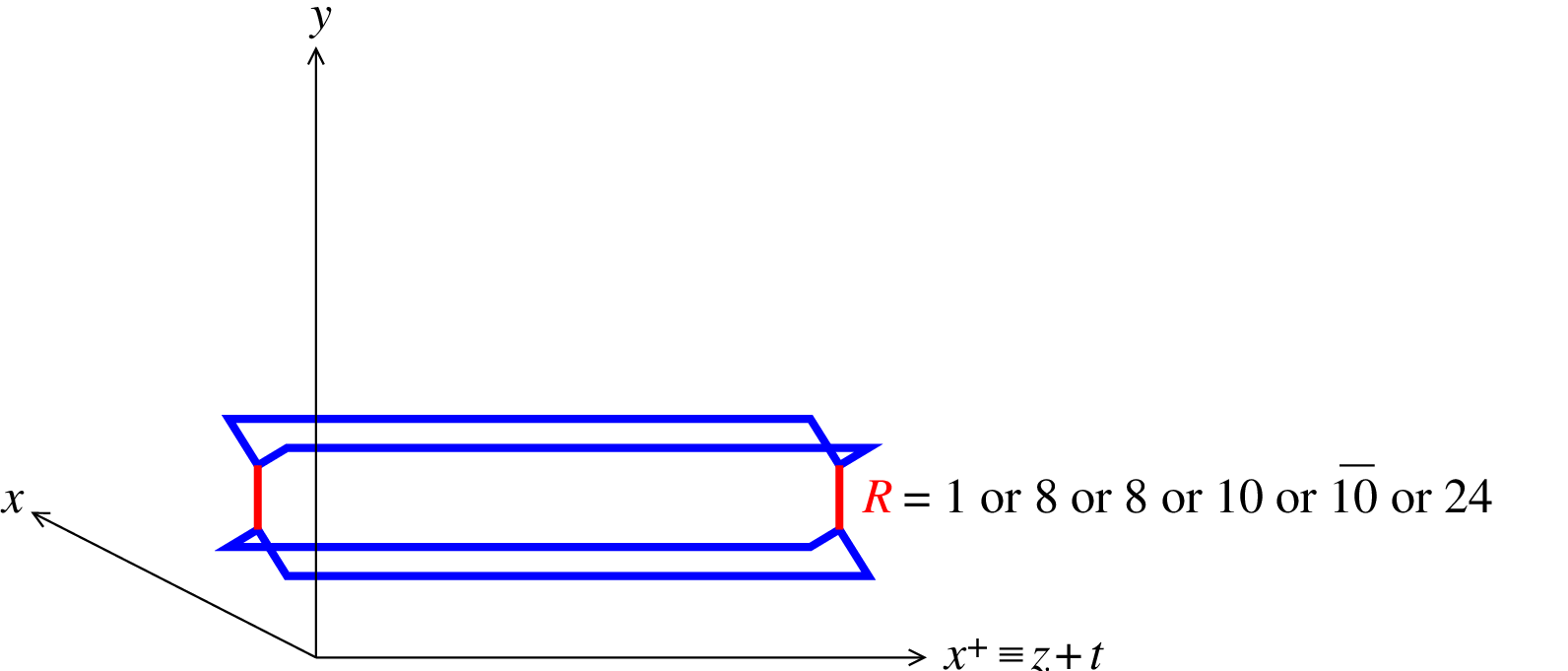}
  \caption{
     \label{fig:Wilson4}
     An example of different ways to contract color for initial and
     final states in the case of an $N$-body potential with $N > 3$.
     The particular example above is for $N{=}4$ light-like
     adjoint Wilson lines.  Here, blue denotes
     adjoint representation.  The two red line segments, however,
     may each be independently chosen to be in any irreducible representation
     $8\otimes8 = 1\oplus8\oplus8\oplus10\oplus\overline{10}\oplus27$.
     Because of the hierarchy of scales discussed in the text, the
     details of the lengths of the red lines, or the transverse positions
     of their endpoints,
     are not relevant to defining
     the potential $\mtrx{V}$
     in the limit of small transverse separations because
     of the hierarchy of scales discussed in the text.
  }
\end {center}
\end {figure}

It is this hierarchy of time scales that makes it possible to consider
color dynamics in the context of a potential approximation $V$.
To define a potential, one needs Wilson loops that are long compared
to the correlation length of the medium ($L \gg \xi$).  But since
the color dynamics scale (\ref{eq:tdecohere}) is {\it also} long
compared to the correlation length, it is then possible to treat that
color dynamics in terms of a potential.  One can imagine picking the initial
and final color states by an appropriate contraction of the initial
and final Wilson lines, as depicted, for example, in
fig.\ \ref{fig:Wilson4}.  Repeating this for all possible choices
in some basis of possible color combinations would allow one to
interpret the Wilson loops of fig.\ \ref{fig:Wilson4} as giving
a matrix result in color representation space,
that can then be written in exponential form $e^{-i \mtrx{V} L}$
to extract a corresponding
matrix result for the potential $\mtrx{V}$ in that basis.
We do not need to do any of this in practice, however, to obtain the
result (\ref{eq:VN}) of this paper.  The point is just to understand
why, for small transverse separations,
it can be sensible to talk about the color structure of $\mtrx{V}$
in the first place.

\subsubsection {Splitting contributions}
\label {sec:dblLog}

\begin {figure}[t]
\begin {center}
  \includegraphics[scale=0.5]{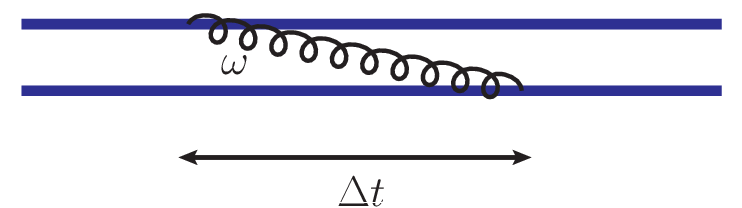}
  \caption{
     \label{fig:V2dblLog}
     A contribution where two light-like Wilson lines are connected
     by a high-energy ($\omega \gg T$), nearly-collinear gluon line.
     Though not drawn explicitly above, the high-energy Wilson
     lines and gluon are all interacting repeatedly with the background
     fields of the plasma.
  }
\end {center}
\end {figure}

The above discussion cheated somewhat, making an implicit assumption
that ignored an additional complication:
I assumed that the only way that the light-like
Wilson lines interact with the plasma is by {\it directly} experiencing
fields already present in the plasma.
However, there are also important contributions \cite{Wu0} where, instead,
a light-like Wilson line emits a {\it high-energy} ($\omega \gg T$)
nearly-collinear gluon, which propagates a long time $\Delta t \gg \xi$
(scattering from the medium the
whole time) before re-attaching to a Wilson line.  An example
is shown in fig.\ \ref{fig:V2dblLog}.
As I'll review, these processes can either (i) be absorbed into the
potential $V$, or else (ii) are suppressed by the running coupling
$\alphas(\mu)$ evaluated at momentum scale $\mu\sim 1/\Delta b$.
Even if the plasma is strongly-coupled, $\alphas(1/\Delta b)$ will be
small for small enough $\Delta b$ (which, formally at least, is the
relevant limit for applications to LPM splitting rates at
large enough energy).

It was shown by Liou, Mueller and Wu \cite{Wu0} that processes
like fig.\ \ref{fig:V2dblLog}
generate corrections $\delta\hat q$ to $\hat q$ of size
$\alphas\hat q$ times a large double logarithm.  In the context
of the 2-body potential, this corresponds to corrections
$\delta V$ to the naive harmonic
potential (\ref{eq:V2intro}) of size%
\footnote{
  The double log shown in (\ref{eq:dV2})
  assumes that the length $L$ of the Wilson loop
  is taken to infinity for fixed $\Delta b$.
  A more general parametric estimate would be to replace the argument
  $1/\hat q\tau_0(\Delta b)^2$ of the log in (\ref{eq:dV2})
  by $\min(t_{\rm decohere},L)/\tau_0$.
  For $\Delta b \to 0$ with fixed $L$, one then recovers the
  $\ln^2(L/\tau_0)$ in eq.\ (45) of ref. \cite{Wu0}, where my
  $\tau_0$ is their $l_0$.
}
\begin {equation}
   \delta V(\Delta b) \sim
   \alphas\hat q \, (\Delta b)^2
       \ln^2\Bigl( \frac{t_{\rm decohere}}{\tau_0} \Bigr)
   \sim
   \alphas\hat q \, (\Delta b)^2
       \ln^2\Bigl( \frac{1}{\hat q\tau_0(\Delta b)^2} \Bigr) ,
\label {eq:dV2}
\end {equation}
where $\tau_0$ is a characteristic scale of the medium that,
for weakly-coupled quark-gluon
plasmas, corresponds to the mean free path for elastic scattering.
Various authors \cite{Blaizot,Iancu,Wu}
have shown that the double-log correction to
$\hat q$ is universal in the sense that, if one computes similar
double-log corrections to QCD LPM splitting rates
(related to 3-body potentials like in fig.\ \ref{fig:Wilson3}),
then those corrections are also completely accounted for by
the correction to $\hat q$ originally found by Liou et al.
They were also able to sum leading logarithms at all orders
in $\alphas$, but the points
I need to make here can already be discussed in the simpler context of
(\ref{eq:dV2}).

Now consider what effect these double-log contributions have
on my argument about the slow rate of mixing of different color
representations for $\mtrx{\Tgen}_1{+}\mtrx{\Tgen}_2$ for $N$-body potentials.
The logarithms in (\ref{eq:dV2}) arise from
time separations $\Delta t$ in fig.\ \ref{fig:V2dblLog} over the
parametrically large range
\begin {subequations}
\label {eq:ranges}
\begin {equation}
   \tau_0 \ll \Delta t \ll t_{\rm decohere}
\label {eq:dtrange}
\end {equation}
in concert with gluon energies $\omega$ in the range
\begin {equation}
  \hat q(\Delta t)^2 \ll \omega \ll (\Delta t)/(\Delta b)^2 ,
\end {equation}
\end {subequations}
[which together restrict $\omega$ to
$T \ll \omega \ll (\Delta t)/(\Delta b)^2$,
where I've interpreted the plasma scale $\hat q \tau_0^2$ as order $T$].%
\footnote{
  I've assumed that $L$ is large compared to all other distance scales in
  (\ref{eq:ranges}).
  For readers who may find these ranges
  more familiar or understandable in terms of other variables,
  here is a translation.
  (Translation 1) I've assumed $\Delta b$ non-zero and
  $L \gg t_{\rm decohere}$ in my analysis, but many analyses instead
  study the case of $\Delta b \to 0$ with $L$ fixed,
  for which $t_{\rm decohere} \gg L$.  One can get the
  parametric ranges in that case from the cross-over case
  $t_{\rm decohere} \sim L$.  From (\ref{eq:tdecohere}), the cross-over
  case corresponds to $\Delta b \sim (\hat q L)^{-1/2}$ and so (\ref{eq:ranges})
  becomes $\tau_0 \ll \Delta t \ll L$ and
  $\hat q (\Delta t)^2 \ll \omega \ll \hat q L \, \Delta t$,
  which together restrict $T \ll \omega \ll \omega_{\rm c} \equiv \hat q L^2$.
  (Translation 2) If interested instead in the application to LPM
  bremsstrahlung of a gluon with energy $\Omega$
  in an arbitrarily {\it thick} plasma ($L \to \infty$),
  one may again use the first translation but replacing $L$ by
  the formation time for the underlying bremsstrahlung, which is of order
  $t_{\rm form}(\Omega) \sim \sqrt{\Omega/\hat q}$.
  That's equivalent to setting $t_{\rm decohere} \sim t_{\rm form}(\Omega)$.
  The typical size of $\Delta b$ for the underlying bremsstrahlung is then
  of order $\Delta b \sim (\hat q t_{\rm form})^{-1/2} \sim (\hat q \Omega)^{-1/4}$.
  The ranges (\ref{eq:ranges})
  [now representing
   double-log effects from additional bremsstrahlung (real or virtual)
   of a second gluon ($\omega$) from the original gluon ($\Omega$)] are
  $\tau_0 \ll \Delta t \ll t_{\rm form}(\Omega)$ and
  $\hat q (\Delta t)^2 \ll \omega \ll \hat q t_{\rm form}(\Omega) \, \Delta t$,
  which together restrict $T \ll \omega \ll \Omega$.
}
Because $\Delta t \ll t_{\rm decohere}$, that means that the
gluon exchange shown in fig.\ \ref{fig:V2dblLog} is {\it local} compared
to the time $t_{\rm decohere}$, at least as far as
log-enhanced effects are concerned.  Earlier, when only considering
direct interactions of Wilson lines with plasma fields, I asserted that
making sense of the color representation of $\mtrx{\Tgen}_1{+}\mtrx{\Tgen}_2$
required the hierarchy of scales $t_{\rm decohere} \gg \xi$ so that
the rate of color change was slow compared to the time scales of the
physics generating the potential $V$. 
But the same argument can still be made for double-log contributions
like fig.\ \ref{fig:V2dblLog}
because we have $t_{\rm decohere} \gg \Delta t$.

One might next be concerned about effects from fig.\ \ref{fig:V2dblLog}
that are {\it not}\/ log enhanced, specifically the contribution from
the upper endpoint $\Delta t \sim t_{\rm decohere}$ of the range
(\ref{eq:dtrange}).  In this case, there is no separation of scale.
The corresponding contribution to the potential is of order
(\ref{eq:dV2}) without the double log factor,
\begin {equation}
   \delta V(\Delta b) \sim
   \alphas\hat q \, (\Delta b)^2 ,
\label {eq:dV2nolog}
\end {equation}
which is suppressed
by an uncompensated factor of $\alphas$
compared to the potential (\ref{eq:V2intro}).
That would not really be a ``suppression''
if the scale for that $\alphas$ were a plasma
scale, e.g.\ $\alphas(\mD)$, since the whole point is to be able
to discuss strongly-coupled plasmas.  However, the
explicit $\alphas$ in (\ref{eq:dV2nolog}) arises from the two
factors of $g$ associated with
where the high-energy gluon connects to the Wilson lines in
fig.\ \ref{fig:V2dblLog}.
The relevant distance scale for this coupling is the typical
relative separation $\Delta B_\perp$ of the gluon from the Wilson lines in
fig.\ \ref{fig:V2dblLog}, which can be estimated from free
gluon diffusion%
\footnote{
  See the discussion of eq.\ (13) of Liou et al.\ \cite{Wu0}.
}
as $\Delta B_\perp \sim \sqrt{(\Delta t)/\omega}$.  In the limiting case
$\Delta t \sim t_{\rm decohere}$ now being discussed, this is
$\Delta B_\perp \sim \Delta b$.
So, in the limit of small $\Delta b$,
the troublesome $\Delta t\sim t_{\rm decohere}$ corrections to the potential
are proportional to $\alphas(1/\Delta b)$ and so are indeed suppressed for
small enough $\Delta b$.  A recent discussion of this scale for $\alpha$ in the
application to overlapping formation times beyond
double-log approximation (a discussion related to the 4-body potential)
can be found in ref.\ \cite{QEDnf}.



\section {Conclusion}
\label {sec:conclusion}

The result of this paper (\ref{eq:VN}) provides the equivalent of the
$\hat q$ approximation but for $N$-body potentials defined with
parallel, light-like Wilson lines.  Such potentials will be
needed, for example, to study
overlapping formation times during in-medium shower development
without taking the large-$\Nc$ limit.
The core argument, given in section \ref{sec:general},
is short and simple; the rest of this paper just frames
it with the necessary background.
In particular, the $N$-body potential has
non-trivial color structure for $N>3$.
A discussion of how this structure could be implemented in calculations
of overlapping formation times for 3-color (as opposed to large-$\Nc$)
QCD requires additional machinery and is left for later
work \cite{color}.


\begin{acknowledgments}

I am greatly indebted to Han-Chih Chang,
Simon Caron-Huot, Guy D. Moore, Hong Liu, Gunnar Bali,
and Yacine Mehtar-Tani for valuable discussions.
This work was supported, in part, by the U.S. Department
of Energy under Grant No.~DE-SC0007984.

\end{acknowledgments}

\appendix

\section{2-point gauge correlators and \boldmath$\Gamma_{\rm el}(b)$}

In this appendix, I review how 2-point correlators between the two
light-like Wilson lines in fig.\ \ref{fig:Wilson2weak} are given by
the $\Gamma_{\rm el}(\b)$ of (\ref{eq:Gammab}).  Many of the original
discussions of the physics of $\hat q$ and QCD LPM splitting rates in
the literature assume particular models for interactions with the
medium such as Debye-screened Coulomb scatterings from static rather
than dynamic scattering centers, ignoring the dependence of
scattering cross-sections on the momentum (and so momentum distribution) of
plasma particles, and/or assuming that individual momentum exchanges
$q$ are soft compared to the plasma temperature $T$.
But I will instead keep the discussion here general.

Let $x$ be a point on one light-like Wilson line and $y$ on the other.
First note that these two points are then space-like separated,
which means (importantly)
that we will not need to worry about time-ordering
of the gauge fields $A(x)$ and $A(y)$ in what follows.
Now use translation invariance to fix $y$ at the origin, and integrate
$x$ along its Wilson line
$(x^0,x^1,x^2,x^3) = (t,\b,t)$ to give the correlator
\begin {equation}
  {\cal C} \equiv
  g^2 \int dt \> \langle v\cdot A(t,\b,t) \, v\cdot A(0) \rangle ,
\end {equation}
where
\begin {equation}
  v^\mu \equiv (1,0,0,1) .
\end {equation}
Rewriting $\langle v\cdot A(x) \, v\cdot A(0) \rangle$ in terms of
its Fourier transform and then performing the $dt$ integral above gives%
\begin {equation}
   {\cal C} =
   g^2 \int_q e^{i\q_\perp\cdot\b}
   \langle v\cdot\tilde A(q)^* \, v\cdot\tilde A(q) \rangle
   \,2\pi\,\delta(v\cdot q) ,
\end {equation}
where the integral is over 4-momentum $q$.
Let $|{\rm i}\rangle$ be any possible state of the medium.
The above is then the appropriate medium-state average of%
\footnote{
  For example, for thermal equilibrium, take $|{\rm i}\rangle$ to
  be exact QFT energy eigenstates and average (\ref{eq:i}) with
  relative weight $e^{-\beta E_{\rm i}}$.  One may do something similar with
  any density matrix describing the medium by working in the basis
  where the density matrix is diagonal.
}
\begin {equation}
   {\cal C} =
   g^2 \int_q e^{i\q_\perp\cdot\b}
   \langle {\rm i} | v\cdot\tilde A(q)^* \, v\cdot\tilde A(q) | {\rm i} \rangle
   \,2\pi\,\delta(v\cdot q) .
\label {eq:i}
\end {equation}
Inserting a complete set of intermediate states $|{\rm f}\rangle$ in the
middle,
\begin {equation}
   {\cal C} =
   \int_q e^{i\q_\perp\cdot\b} \sum_{\rm f}
   \bigl|\langle {\rm f} | g v\cdot\tilde A(q) | {\rm i} \rangle \bigr|^2
   \,2\pi\,\delta(v\cdot q) .
\label {eq:Cfinal}
\end {equation}

Now consider instead calculating the elastic scattering rate of a
particle with very high energy $E$ via exchanging a gluon with the medium.
In the limit that $E$ is much higher than the exchanged momentum $q$,
the particle-gluon vertex is $i gv\cdot A$ times a relativistic normalization
factor of $2E$.  The rate is then
\begin {equation}
   \Gamma_{\rm el} =
   \sum_{\rm f} \int_q \frac{1}{2E} \,
   \bigl|\langle {\rm f} | g v\cdot\tilde A(q) \, 2E| {\rm i} \rangle \bigr|^2
   \,2\pi\,\delta\bigl((P+q)^2\bigr) ,
\end {equation}
where $P \equiv (E,0,0,E)$ is the high-energy particle's 4-momentum,
$1/2E$ is the usual initial-state normalization factor, and
$\delta\bigl((P+q)^2\bigr)$ puts the final state of the high-energy
particle on shell.  Remember that the states $|{\rm i}\rangle$ and
$|{\rm f}\rangle$ above refer to states of the medium and do not
include the states of the high-energy particle, which here have been
treated explicitly.  Making use of the high-energy limit for the $P$
inside the $\delta$-function,
\begin {equation}
   \Gamma_{\rm el} =
   \sum_{\rm f} \int_q
   \bigl|\langle {\rm f} | g v\cdot\tilde A(q) | {\rm i} \rangle \bigr|^2
   \,2\pi\,\delta(v\cdot q) .
\end {equation}
Dropping the $\q_\perp$ integration above,
\begin {equation}
   \frac{d\Gamma_{\rm el}}{d^2q_\perp} =
   \sum_{\rm f} \int_{q^0,q^z}
   \bigl|\langle {\rm f} | g v\cdot\tilde A(q) | {\rm i} \rangle \bigr|^2
   \,2\pi\,\delta(v\cdot q) .
\end {equation}
Plugging this into the definition (\ref{eq:Gammab}) of $\Gamma_{\rm el}(\b)$
then shows that that the correlator ${\cal C}$ (\ref{eq:Cfinal}) between
two Wilson lines is the same as $\Gamma_{\rm el}(\b)$.



\begin{thebibliography}{}

\bibitem{LRW1}
  H.~Liu, K.~Rajagopal and U.~A.~Wiedemann,
  ``Calculating the jet quenching parameter from AdS/CFT,''
  Phys.\ Rev.\ Lett.\  {\bf 97}, 182301 (2006)
  [hep-ph/0605178].

\bibitem{LRW2}
  H.~Liu, K.~Rajagopal and U.~A.~Wiedemann,
  ``Wilson loops in heavy ion collisions and their calculation in AdS/CFT,''
  JHEP {\bf 0703}, 066 (2007)
  [hep-ph/0612168];

\bibitem{SimonNLO}
  S.~Caron-Huot,
  ``O(g) plasma effects in jet quenching,''
  Phys.\ Rev.\ D {\bf 79}, 065039 (2009)
  [arXiv:0811.1603 [hep-ph]].

\bibitem{Casimir1}
  C.~Anzai, Y.~Kiyo and Y.~Sumino,
  ``Violation of Casimir Scaling for Static QCD Potential at Three-loop Order,''
  Nucl.\ Phys.\ B {\bf 838}, 28 (2010)
  Erratum: [Nucl.\ Phys.\ B {\bf 890}, 569 (2015)]
  [arXiv:1004.1562 [hep-ph]].

\bibitem{Casimir2}
  R.~N.~Lee, A.~V.~Smirnov, V.~A.~Smirnov and M.~Steinhauser,
  ``Analytic three-loop static potential,''
  Phys.\ Rev.\ D {\bf 94}, no. 5, 054029 (2016)
  [arXiv:1608.02603 [hep-ph]].

\bibitem{Casimir3}
  A.~Grozin, J.~Henn and M.~Stahlhofen,
  ``On the Casimir scaling violation in the cusp anomalous dimension
    at small angle,''
  JHEP {\bf 1710}, 052 (2017)
  [arXiv:1708.01221 [hep-ph]].

\bibitem{BDMPS12}
  R.~Baier, Y.~L.~Dokshitzer, A.~H.~Mueller, S.~Peigne and D.~Schiff,
  ``The Landau-Pomeranchuk-Migdal effect in QED,''
  Nucl.\ Phys.\  B {\bf 478}, 577 (1996)
  [arXiv:hep-ph/9604327];
  ``Radiative energy loss of high-energy quarks and gluons in a
    finite volume quark - gluon plasma,''
  {\it ibid.}\ {\bf 483}, 291 (1997) [arXiv:hep-ph/9607355].

\bibitem{BDMPS3}
  R.~Baier, Y.~L.~Dokshitzer, A.~H.~Mueller, S.~Peigne and D.~Schiff,
  ``Radiative energy loss and $p_\perp$-broadening of high energy partons in
    nuclei,''
  {\it ibid.}\ {\bf 484} (1997)
  [arXiv:hep-ph/9608322].

\bibitem{Zakharov}
 B.~G.~Zakharov,
 ``Fully quantum treatment of the Landau-Pomeranchuk-Migdal effect in
   QED and QCD,''
 JETP Lett.\  {\bf 63}, 952 (1996)
 [Pis'ma Zh.\ \'Eksp.\ Teor.\ Fiz.\  {\bf 63}, 906 (1996)]
 [arXiv:hep-ph/9607440];
 ``Radiative energy loss of high-energy quarks in finite size nuclear matter an
   quark - gluon plasma,''
 {\it ibid.}\  {\bf 65}, 615 (1997)
 [Pis'ma Zh.\ \'Eksp.\ Teor.\ Fiz.\  {\bf 65}, 585 (1997)]
 [arXiv:hep-ph/9607440].

\bibitem{2brem}
  P.~Arnold and S.~Iqbal,
  ``The LPM effect in sequential bremsstrahlung,''
  JHEP {\bf 04}, 070 (2015)
  [{\it erratum} JHEP {\bf 09}, 072 (2016)]
  [arXiv:1501.04964 [hep-ph]].

\bibitem{Blaizot}
  J.~P.~Blaizot and Y.~Mehtar-Tani,
  ``Renormalization of the jet-quenching parameter,''
  Nucl.\ Phys.\ A {\bf 929}, 202 (2014)
  [arXiv:1403.2323 [hep-ph]].

\bibitem{Iancu}
  E.~Iancu,
  ``The non-linear evolution of jet quenching,''
  JHEP {\bf 1410}, 95 (2014)
  [arXiv:1403.1996 [hep-ph]].

\bibitem{Wu}
  B.~Wu,
  ``Radiative energy loss and radiative $p_{\bot}$-broadening of
    high-energy partons in QCD matter,''
  JHEP {\bf 1412}, 081 (2014)
  [arXiv:1408.5459 [hep-ph]].

\bibitem{seq}
  P.~Arnold, H.~C.~Chang and S.~Iqbal,
  ``The LPM effect in sequential bremsstrahlung 2: factorization,''
  JHEP {\bf 1609}, 078 (2016)
  [arXiv:1605.07624 [hep-ph]];

\bibitem{dimreg}
  P.~Arnold, H.~C.~Chang and S.~Iqbal,
  ``The LPM effect in sequential bremsstrahlung: dimensional regularization,''
  JHEP {\bf 1610}, 100 (2016)
  [arXiv:1606.08853 [hep-ph]].

\bibitem{4point}
  P.~Arnold, H.~C.~Chang and S.~Iqbal,
  ``The LPM effect in sequential bremsstrahlung: 4-gluon vertices,''
  JHEP {\bf 1610}, 124 (2016)
  [arXiv:1608.05718 [hep-ph]].

\bibitem{simple}
  P.~B.~Arnold,
  ``Simple Formula for High-Energy Gluon Bremsstrahlung
    in a Finite, Expanding Medium,''
  Phys.\ Rev.\ D {\bf 79}, 065025 (2009)
  [arXiv:0808.2767 [hep-ph]].

\bibitem{ArnoldDogan}
  P.~B.~Arnold and C.~Dogan,
  ``QCD Splitting/Joining Functions at Finite Temperature
    in the Deep LPM Regime,''
  Phys.\ Rev.\ D {\bf 78}, 065008 (2008)
  [arXiv:0804.3359 [hep-ph]].

\bibitem{Peshier}
  A.~Peshier,
  ``QCD running coupling and collisional jet quenching,''
  J.\ Phys.\ G {\bf 35}, 044028 (2008).

\bibitem{Wu0}
  T.~Liou, A.~H.~Mueller and B.~Wu,
  ``Radiative $p_\bot$-broadening of high-energy quarks and gluons in
    QCD matter,''
  Nucl.\ Phys.\ A {\bf 916}, 102 (2013)
  [arXiv:1304.7677 [hep-ph]].

\bibitem{ZakharovHO}
  B.~G.~Zakharov,
  ``On the energy loss of high-energy quarks in a finite size
    quark-gluon plasma,''
  JETP Lett.\  {\bf 73}, 49 (2001)
  [Pisma Zh.\ Eksp.\ Teor.\ Fiz.\  {\bf 73}, 55 (2001)]
  [hep-ph/0012360].

\bibitem{HO}
  P.~B.~Arnold,
  ``High-energy gluon bremsstrahlung in a finite medium:
    harmonic oscillator versus single scattering approximation,''
  Phys.\ Rev.\ D {\bf 80}, 025004 (2009)
  doi:10.1103/PhysRevD.80.025004
  [arXiv:0903.1081 [nucl-th]].

\bibitem{D'Eramo}
  F.~D'Eramo, H.~Liu and K.~Rajagopal,
  ``Transverse Momentum Broadening and the Jet Quenching Parameter, Redux,''
  Phys.\ Rev.\ D {\bf 84}, 065015 (2011)
  [arXiv:1006.1367 [hep-ph]].

\bibitem{Benzke}
  M.~Benzke, N.~Brambilla, M.~A.~Escobedo and A.~Vairo,
  ``Gauge invariant definition of the jet quenching parameter,''
  JHEP {\bf 1302}, 129 (2013)
  [arXiv:1208.4253 [hep-ph]].

\bibitem{LRW3}
  H.~Liu, K.~Rajagopal and Y.~Shi,
  ``Robustness and Infrared Sensitivity of Various Observables
    in the Application of AdS/CFT to Heavy Ion Collisions,''
  JHEP {\bf 0808}, 048 (2008)
  doi:10.1088/1126-6708/2008/08/048
  [arXiv:0803.3214 [hep-ph]].

\bibitem{YacineColorSplit}
  Y.~Mehtar-Tani and K.~Tywoniuk,
  ``Jet coherence in QCD media: the antenna radiation spectrum,''
  JHEP {\bf 1301}, 031 (2013)
  [arXiv:1105.1346 [hep-ph]].

\bibitem{QEDnf}
  P.~Arnold, S.~Iqbal and T.~Rase,
  ``Strong- vs. weak-coupling pictures of jet quenching: a dry run using QED,''
  arXiv:1810.06578 [hep-ph].

\bibitem{color}
  P. Arnold and H. C. Chang,
  in preparation.

\end{thebibliography}
\end {document}